\documentclass[preprint,showpacs,preprintnumbers,superscriptaddress,article]{revtex4-1}
\usepackage{multirow}
\usepackage{diagbox}
\usepackage{booktabs}
\usepackage{array}
\usepackage{natbib}
\usepackage{amsmath,amssymb,graphicx,bm,subfigure,float,siunitx,color}
\usepackage{bm,subfigure,float,siunitx,graphicx}
\usepackage[utf8]{inputenc}
\usepackage[figuresright]{rotating}
\usepackage{color}
\usepackage{epstopdf}
\usepackage{mathrsfs}
\usepackage{amsmath}

\newcommand{\be}{\begin{equation}}
\newcommand{\ee}{\end{equation}}

\newcommand{\1}{\left}
\newcommand{\2}{\right}
\def\({\left(}
\def\){\right)}
\def\[{\left[}
\def\]{\right]}

\newcommand{\dif}{\,\mathrm{d}}

\newcommand{\me}{\mathrm{e}}

\newcommand{\p}{\partial}

\newcommand{\m}{\mu}
\newcommand{\n}{\nu}
\newcommand{\al}{\alpha}

\renewcommand{\th}{\theta}

\newcommand{\na}{\nabla}

\begin{document}
\title{Thin accretion disk and shadow of Kerr-Sen black hole in Einstein-Maxwell-dilaton-axion gravity}
\author{Haiyuan Feng}
\email{Email address:  fenghaiyuanphysics@gmail.com}
\affiliation{Department of Physics, Southern University of Science and Technology, Shenzhen 518055, Guangdong, China}

\author{Rong-Jia Yang \footnote{Corresponding author}}
\email{Email address: yangrongjia@tsinghua.org.cn}
\affiliation{College of Physical Science and Technology, Hebei University, Baoding 071002, China}

\author{Wei-Qiang Chen\footnote{Corresponding author}}
\email{Email address: chenwq@sustech.edu.cn}
\affiliation{Department of Physics, Southern University of Science and Technology, Shenzhen 518055, Guangdong, China}

\begin{abstract}

We investigate the thin disk and shadow of Kerr-Sen black hole in Einstein-Maxwell-dilaton-axion gravity. The results reveal that as the dilaton parameter $r_2$ increase, the energy flux, the radiation temperature, the spectra luminosity, and the radiative efficiency of the disk all increase. By narrowing down the dilaton parameter range to  $0\leqslant \frac{r_2}{M}\leqslant0.4$, we discover that in the high-frequency region, the Kerr-Sen black hole demonstrates higher energy output compared to the Kerr black hole. We also investigated the shadow of Kerr-Sen black hole in a uniform plasma environment. For fixed inclination angle, dilaton, and spin parameters, the shadow increases as the homogeneous plasma parameter $k$ increases. Conversely, when $k $ and $a$ are fixed, an increase in $r_2$ leads to a decrease in the shadow. Finally, we constrain the model parameters with observational data from M87* and Sgr A*.

\end{abstract}

\maketitle

\section{Introduction}
Black holes (BHs) predicted by general relativity (GR) are intriguing celestial entities and the most potent sources of gravitational fields in cosmology. They often exhibit high spin and strong magnetic fields, making them ideal for studying material properties and gravitational effects. Recent observations have significantly supported the existence of BHs. A major breakthrough was the detection of gravitational waves from merging binary BHs by the LIGO and Virgo teams \cite{LIGOScientific:2016aoc}. Another milestone was the first image of the M87* shadow \cite{EventHorizonTelescope:2019dse,EventHorizonTelescope:2019ths}, followed by the Event Horizon Telescope (EHT) capturing an image of the supermassive BH at the center of our galaxy, Sgr A*\cite{EventHorizonTelescope:2022wkp}. Additionally, BHs are confirmed through the electromagnetic spectrum of accretion disks \cite{Nampalliwar:2018tup,book,Yuan:2014gma}. These achievements deepen our understanding of the characteristics of accretion disks near the event horizon of supermassive compact objects.

Compact objects like BHs gain mass through accretion, sustained by the presence of an accretion disk. An accretion disk is a structure of diffuse material, such as gas and dust, that spirals inward towards the central object due to gravitational forces, releasing gravitational energy as radiation. This radiation can be observed and analyzed, providing valuable information about the properties and processes near these compact objects. Images of accretion disks around BHs are a focus of observational astronomy. The standard model of geometrically thin and optically thick accretion disks was first proposed by Shakura and Sunyaev and later extended to include GR by Page, Novikov, and Thorne \cite{Shakura:1972te,1973blho.conf..343N,Page:1974he,Thorne:1974ve}. They assumed a steady-state accretion disk with a constant mass accretion rate $\dot{M}$, in hydrodynamic and thermodynamic equilibrium, ensuring a black body electromagnetic spectrum. This model successfully explains the observed spectral features of astrophysical BHs.

On the other hand, the study of BH's shadow in a plasma environment is an emerging field that bridges astrophysics and GR, enhancing our understanding of BHs and the interstellar medium. Plasma not only influences light trajectories through the gravitational effects of compact objects but also acts as a dispersive medium \cite{perlick2000ray,EventHorizonTelescope:2019dse}. Thus, it is reasonable to assume that BHs and other exotic compact objects are surrounded by plasma in realistic astrophysical environments. Investigating the effects of plasma on BH's shadows allows us to probe the interactions between radiation and plasma near BHs, offering insights into the plasma's density and distribution. This research refines theoretical models of BH physics, as plasma's influence must be accounted for in precise measurements of BH's parameters. As observational techniques advance, especially with high-resolution instruments like the EHT, understanding plasma's role becomes increasingly crucial for accurate interpretation of BH's shadows.

Ultimately, the discernible features in the emission spectra of accretion disks and the observation of BH's shadows provide valuable insights into the characteristics of the central massive object. These observations also test modified theories. Extensive research has explored the characteristics of thin disks \cite{Abbas:2023rzk,Karimov:2018whx,Bhattacharyya:2001va,Kovacs:2009gt,Heydari-Fard:2020ugv,Zhang:2021hit,Karimov:2018whx,Kazempour:2022asl,Collodel:2021gxu, He:2022lrc,Shahidi:2020bla,Uniyal:2022vdu,Liu:2024brf,Teodoro:2021ezj, Soroushfar:2020kgb,Schroven:2017jsp,Guo_2024} and BH's shadows in diverse background spacetimes \cite{Rosa:2023qcv,Luminet:1979nyg,PhysRevD.90.062007,Takahashi:2005hy,Takahashi:2005hy,PhysRevD.97.064021,PhysRevLett.115.211102,Li:2013jra,
Amir:2016cen,Cunha:2016wzk,PhysRevD.88.064004,Wang:2017hjl,Uniyal:2022xnq,Abdujabbarov:2015rqa,Okyay:2021nnh,Afrin:2022ztr,Atamurotov:2015nra,He:2022lrc,Harko:2009rp,
Karimov:2018whx,Kazempour:2022asl,Gyulchev:2021dvt,Heydari-Fard:2020ugv,Liu:2020vkh,Pun:2008ua,Zare:2024dtf,Ma:2022jsy,Vagnozzi:2022moj,Mirzaev:2023oud,Ma:2024iao,Xu:2017vse}. Among many alternative theories, the Einstein-Maxwell-dilaton-axion (EMDA) model \cite{Rogatko:2002qe,Sen:1992ua} has attracted significant attention. The accretion onto this BH has been discussed in \cite{Feng:2024mey}. In this model, the dilaton field and the pseudoscalar axion are integrated, both intricately linked to the metric and the Maxwell field. The origins of the dilaton and axion fields can be traced back to string compactifications, presenting compelling implications for inflationary and late-time accelerated cosmologies \cite{Catena:2007jf,Sonner:2006yn}. Consequently, we will focus on the influence of theoretical parameters on the properties of the thin accretion disk and shadow of Kerr-Sen BH in EMDA model.

The dilaton parameter $r_2\equiv\frac{Q^2}{M}$ of Kerr-Sen BH has been constrained through observations in EMDA theory. For example, a preferred value of $r_2\thickapprox0.2M$ was determined based on the optical continuum spectrum of quasars \cite{Banerjee:2020qmi}. Additionally, recent investigations have imposed observational constraints on the parameter as: $0.1M\lesssim r_2\lesssim0.4M$ \cite{Sahoo:2023czj}.  Furthermore, the reference \cite{Tripathi:2021rwb} inferred the constraint of the dilaton parameter a $r_2 < 0.011M$ from the analysis of BH X-ray data. Finally, a constraint on the $r_2$ was obtained by employing simulated data replicating potential observations of the $S2$ star via a gravity interferometer \cite{Fernandez:2023kro}. This demonstrated that enhanced astrometric accuracy can effectively narrow down the acceptable range of the dilaton parameter to $r_2 \lesssim0.066M$.

The article is organized as follows: In Sec. II, the EMDA model and the Kerr-Sen BH will be briefly reviewed. In Sec. III, we will present the geodesic equation governing the motion of timelike particles moving in the equatorial plane. We will plot the specific angular momentum $\tilde{L}$, the specific energy $\tilde{E}$, the angular velocity $\Omega$, and the innermost stable circular orbit $r_{isco}$ of particle with respect to the dilaton parameter $r_2$. We will investigate the physical characteristics of thin accretion disk surrounding Kerr-Sen BH, with a particular focus on the influence of the dilaton parameter on the energy flux, the temperature, and the emission spectrum of the accretion process. In Sec. IV, our main objective is to study the shadow of Kerr-Sen BH within a uniform plasma environment. In section V, we will constrain  the parameter ranges for two BHs using observational data from M87* and Sgr A*. Finally, Sec. VI addresses conclusions and discussions. For convenience, we will use signature convention $(-,+,+,+)$ for the spacetime metric throughout the article.

\section{Kerr-Sen black hole in Einstein-Maxwell-dilaton-axion gravity}
The EMDA model is derived from the low-energy limit behavior of heterotic string theory. It is composed of the dilaton field $\chi$, a gauge vector field $A_{\m}$, the metric $g_{\m\n}$, and a pseudo-scalar axion field $\xi$ \cite{Rogatko:2002qe,Sen:1992ua,Campbell:1992hc}. The action of the EMDA model can be formulated through the coupling of supergravity and super-Yang Mills theory, and it can be described by the
\be
\label{1}
S=\frac{1}{16\pi}\int\sqrt{-g}d^4x\[R-2\p_{\m}\chi\p^{\m}\chi-\frac{1}{2}\me^{4\chi}\p_{\m}\xi\p^\m\xi+
\me^{-2\chi}f_{\m\n}f^{\m\n}+\xi f_{\m\n}\widetilde{f}^{\m\n} \],
\ee
where $R$ represents the Ricci scalar and $f_{\m\n}$ denotes the second-order antisymmetric Maxwell field strength tensor, given by $f_{\m\n}=\nabla_{\m}A_{\n}-\nabla_{\n}A_{\m}$, while $\widetilde{f}^{\m\n}$ signifies the dual tensor of the field strength. The variation of these four fields leads to the following motion equations
\be
\1\{\begin{split}
\label{2}
&\Box\chi-\frac{1}{2}\me^{4\chi}\na_\m\xi\na^\m\xi+\frac{1}{2}\me^{-2\chi}f_{\m\n}f^{\m\n}=0,\\
&\Box\xi+4\na_\m\xi\na^\m\xi-\me^{-4\chi}f_{\m\n}\widetilde{f}^{\m\n}=0,\\
&\na_\m\widetilde{f}^{\m\n}=0,\\
&\na_\m(\me^{-2\chi}f^{\m\n}+\xi\widetilde{f}^{\m\n})=0,\\
&G_{\m\n}=\me^{2\chi}(4f_{\m\rho}f^{\rho}_{\n}-g_{\m\n}f^2)-g_{\m\n}(2\na_\m\chi\na^\m\chi+\frac{1}{2}\me^{4\chi}\na_\m\xi\na^\m\xi)\\
&+\na_\m\chi\na_\n\chi+\me^{4\chi}\na_\m{\xi}\na_\n{\xi}.
\end{split}\2.
\ee
It demonstrates that the dilaton field, axion field, electromagnetic field, and gravitational field are intricately coupled. The axisymmetric classical solution known as the Kerr-Sen BH can be derived in the  Boyer-Lindquist coordinates, and it takes the following form \cite{Garfinkle:1990qj}.
\be
\label{3}
\begin{split}
\dif s^{2}&=-\(1-\frac{2M r}{\tilde{\Sigma}}\)\dif t^{2}+\frac{\tilde{\Sigma}}{\tilde{\Delta}}\dif r^{2}+\tilde{\Sigma}\dif \th^{2}-\frac{4aMr}{\tilde{\Sigma}}\sin^2{\th}\dif t\dif\phi\\
&+\sin^2{\th}\dif \phi^2\[r(r+r_2)+a^2+\frac{2Mra^2\sin^2{\th}}{\tilde{\Sigma}}\],
\end{split}
\ee
with
\be
\1\{\begin{split}
\label{4}
&\tilde{\Sigma}=r(r+r_2)+a^2\cos^2{\th},\\
&\tilde{\Delta}=r(r+r_2)-2Mr+a^2,
\end{split}\2.
\ee
where $M$ represents the mass parameter of the BH, the dilation parameter $r_2=\frac{Q^2}{M}$ is determined by the electric charge $Q$, and $a$ denotes the BH's angular momentum per unit mass. Equation \eqref{3} indicates that upon removing the BH's rotation parameter, a spherically symmetric dilaton BH remains, characterized by mass, electric charge, and asymptotic value \cite{Garfinkle:1990qj}. When the dilaton parameter $r_2$ becomes zero, the Kerr-Sen solution reverts to the Kerr BH.

The BH's event horizon $r_\pm$ is specified by $\tilde{\Delta}=0$, so we can get
\be
\1\{\begin{split}
\label{5}
&r_+=M-\frac{r_2}{2}+\sqrt{\(M-\frac{r_2}{2}\)^2-a^2},\\
&r_-=M-\frac{r_2}{2}-\sqrt{\(M-\frac{r_2}{2}\)^2-a^2}.
\end{split}\2.
\ee
Equation \eqref{5} establishes the theoretical range for the existence of internal and external event horizons as: $0\leqslant\frac{r_2}{M}\leqslant2(1-\frac{a}{M})$ or $-(1-\frac{r_2}{2M})\leqslant\frac{a}{M}\leqslant1-\frac{r_2}{2M}$. As the spin parameter $a$ should not exceed the BH mass $M$, we deduce that the theoretical effective range for $r_2$ is $0\leqslant\frac{r_2}{M}\leqslant2$. Figure 1 depicts the BH region within the allowed parameter range ($r_{+}>r_{-}$), while the white region represents the naked singularity appear ($r_{-}>r_{+}$).
\begin{figure}[H]
\centering
\begin{minipage}{0.5\textwidth}
\centering
\includegraphics[scale=0.7,angle=0]{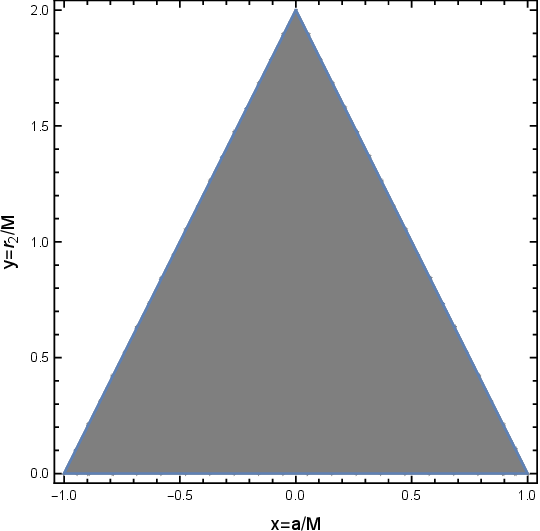}
\end{minipage}%
\caption{\label{Fig.1}{On the horizontal axis, we have the dimensionless BH parameter $a$, while the vertical axis represents the dimensionless dilaton parameter $r_2$. The gray area corresponds to the BH region, whereas the white area denotes the naked singularity.}}
\end{figure}
\section{The physical properties of thin accretion disks}
In this section, we will calculate the radiant energy flux, the radiation temperature, and the observed luminosity from the accretion disc within the permissible interval $0\leqslant \frac{r_2}{M}\leqslant0.4$. This analysis enables the exploration of the observable effects of the Kerr-Sen BH and offers an approach to distinguish it from the Kerr BH. We adopt the fundamental feature of the Novikov-Thorne model to delineate the continuum spectrum  \cite{Novikov:1973kta}. The model incorporates several pivotal assumptions: (1) the mass accretion rate is assumed to be constant and independent of the disk's radius, indicating that the disk is in a steady state; (2) The accreting matter is assumed to exhibit Keplerian motion, which necessitates that the central mass is free from a strong magnetic field; (3) The accretion disk is geometrically thin, with its vertical size negligible compared to its horizontal extent, allowing the heat generated by stress and dynamic friction to be efficiently dissipated through radiation across its surface; (4) The radiation emitted from the disk is treated as black body radiation, resulting from its thermodynamic equilibrium, and is emitted solely perpendicularly to the disk's plane, indicating that the disk is optically thick; (5) the primary contribution to the continuum spectrum originates from electromagnetic emission from the accretion disk surrounding the BH; (6) The spacetime surrounding the central massive object is stationary and axisymmetric, remaining unchanged as it is asymptotically flat and symmetric with respect to reflection across the equatorial plane; (7) The self-gravity of the disk is negligible, meaning that the mass of the disk has no significant effect on the background metric; (8) the particles around the compact central object traverse between the outer edge ($r_{out}$) and the radius of the innermost stable circular orbit ($r_{isco}$) which defined as the inner edge of the disk; (9) the accretion disk is situated in the equatorial plane of the accreting compact object, with the spin of the BH perpendicular to the disk surface.

Novikov-Thorne model in EMDA theory has been discussed in \cite{Banerjee:2020qmi}, here we reconsider this issue in spherical coordinates, which has not been considered before.

\subsection{Timelike geodesics equations }
The formation of an accretion disk is a result of particles moving along geodesic paths in orbits around a compact central object. However, the motion of particles are entirely dependent on the background geometry and the structure of the metric for a stationary axisymmetric spacetime is expressed as
\be
\label{6}
\begin{split}
\dif s^{2}&=g_{tt}\dif t^2+2g_{t\phi}\dif t\dif \phi+g_{rr}\dif r^2+g_{\th\th}\dif \theta^2+g_{\phi\phi}\dif \phi^2,
\end{split}
\ee
in Eq. \eqref{6}, owing to the symmetry of the spacetime, the metric components $g_{tt}$, $g_{t\phi}$, $g_{rr}$, $g_{\th\th}$, and $g_{\phi\phi}$ solely rely on the variables $r$ and $\th$. It is evident that within the aforementioned geometry, the motion exhibits two conserved quantities: the specific energy at infinity denoted as $\tilde{E}$, and the $z$-component of the specific angular momentum $\tilde{L}$ at infinity. These quantities can be obtained by
\be
\1\{\begin{split}
\label{7}
&g_{tt}\frac{\dif t}{\dif \tau}+g_{t\phi}\frac{\dif \phi}{\dif \tau}=-\tilde{E},\\
&g_{t\phi}\frac{\dif t}{\dif \tau}+g_{\phi\phi}\frac{\dif \phi}{\dif \tau}=\tilde{L},
\end{split}\2.
\ee
where $\tau$ denotes the affine parameter. In the equatorial plane characterized by $\th = \frac{\pi}{2}$, the geodesic equation can be derived as follows
\be
\1\{\begin{split}
\label{8}
&\frac{\dif t}{\dif \tau}=\frac{\tilde{E} g_{\phi\phi}+\tilde{L} g_{t\phi}}{g^2_{t\phi}-g_{tt}g_{\phi\phi}},\\
&\frac{\dif \phi}{\dif \tau}=-\frac{\tilde{E} g_{t\phi}+\tilde{L} g_{tt}}{g^2_{t\phi}-g_{tt}g_{\phi\phi}},\\
&g_{rr}\(\frac{\dif r}{\dif \tau}\)^2=V_{eff}(r),
\end{split}\2.
\ee
with the effective potential
\be
\label{9}
V_{eff}=\frac{\tilde{E}^2g_{\phi\phi}+2\tilde{E}\tilde{ L} g_{t\phi}+\tilde{L}^2 g_{tt}}{g^2_{t\phi}-g_{tt}g_{\phi\phi}}-1.
\ee
The effective potential can be analogized to the Newtonian gravitational potential, as it provides the condition under which particles undergo circular orbital motion. The circular orbit within the equatorial plane ($\th=\frac{\pi}{2}$) is defined by the conditions $V_{eff}(r) = 0$ and $V_{eff,r}(r) = 0$. These conditions establish the specific energy $\tilde{E}$ and the specific angular momentum $\tilde{L}$ as functions of the angular velocity $\Omega$ of particles.
\be
\1\{\begin{split}
\label{10}
&\tilde{E}=\frac{-g_{tt}-\Omega g_{t\phi}}{\sqrt{-g_{tt}-2\Omega g_{t\phi}-\Omega^2 g_{\phi\phi}}}\\
&\tilde{L}=\frac{g_{t\phi}+\Omega g_{\phi\phi}}{\sqrt{-g_{tt}-2\Omega g_{t\phi}-\Omega^2 g_{\phi\phi}}},
\end{split}\2.
\ee
where the angular velocity of the test particle $\Omega$ is
\be
\label{11}
\Omega=\frac{\dif \phi}{\dif t}=\frac{  -g_{t\phi,r} \pm \sqrt{(-g_{t\phi,r})^2-g_{\phi\phi,r} g_{tt,r} }       }{g_{\phi\phi,r}}.
\ee
Furthermore, to ascertain the inner edge of the disk, it is imperative to identify the innermost stable circular orbit (ISCO) of the BH potential \cite{Kovacs:2010xm}. This involves utilizing the condition $V_{eff,rr}(r) = 0$, which leads to the derivation of the following relation.
\be
\label{12}
\tilde{E}^2 g_{\phi\phi,rr}+2\tilde{E}\tilde{ L} g_{t\phi,rr}+\tilde{L}^2 g_{tt,rr}-(g^2_{t\phi}-g_{tt}g_{\phi\phi})_{,rr}=0.
\ee
For $r <r_{isco}$, the equatorial circular orbits are unstable, so $r_{isco}$ determines the inner edge of the thin accretion disk \cite{Shahidi:2020bla}. In Fig. 2, we illustrate the variations of particle's angular velocity, specific energy, specific angular momentum and $r_{isco}$ with respect to the distance $r$ and the spin parameter $a$.

Based on Fig. 2, we see that the angular velocity of the Kerr BH surpasses that of the Kerr-Sen BH under for a fixed distance $r$. As the distance $r$ increases, the angular velocity exhibits a decreasing trend. However, the specific energy $\tilde{E}$ and the specific angular momentum $\tilde{L}$ demonstrate distinct patterns. It is worth noting that both $\tilde{E}$ and $\tilde{L}$ exhibit higher magnitudes in GR compared to the EMDA theory. Furthermore, as the parameter $r_2$ increases, the deviation from the Kerr BH notably diminishes, eventually converging. Additionally, from the last diagram, we obtain that within the allowed parameter range, as the parameter $a$ increases, the innermost stable circular orbit gradually decreases for different $r_2$.
\begin{figure}[H]
\centering
\begin{minipage}{0.48\textwidth}
\centering
\includegraphics[scale=0.78,angle=0]{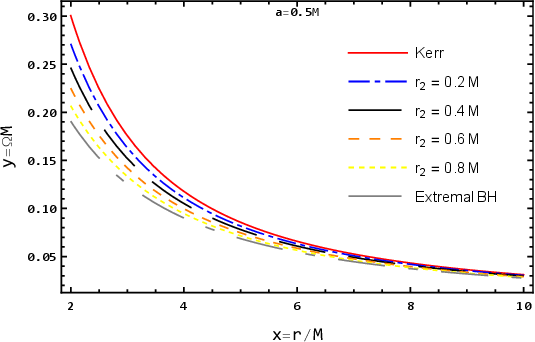}
\end{minipage}%
\begin{minipage}{0.5\textwidth}
\centering
\includegraphics[scale=0.8,angle=0]{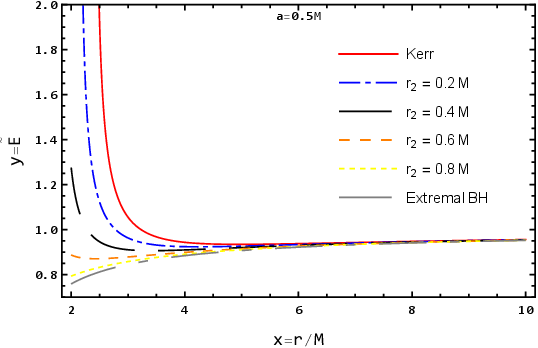}
\end{minipage}
\begin{minipage}{0.49\textwidth}
\includegraphics[scale=0.78,angle=0]{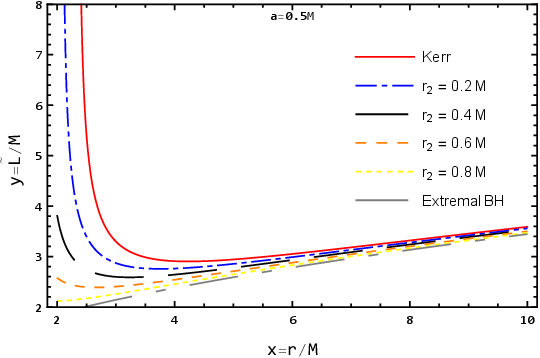}
\end{minipage}%
\begin{minipage}{0.45\textwidth}
\centering
\includegraphics[scale=0.78,angle=0]{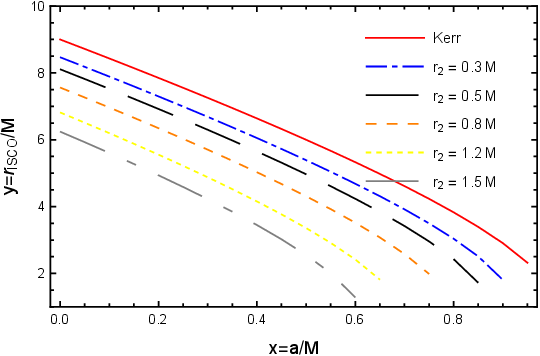}
\end{minipage}
\caption{\label{Fig.2} The angular velocity $\Omega$ (top-left panel), the specific energy $\tilde{E}$ (top-right panel), and the specific angular momentum $\tilde{L}$ (bottom-left panel) are shown as a function of the radial coordinate $r$ for different values of $r_2$ with $\frac{a}{M}=0.5$. In bottom-right panel, $r_{isco}$ illustrates the variation of $a$ under different dilaton parameters $r_2$. }
\end{figure}

\subsection{Continuum Spectrum in the Kerr-Sen black hole}
In model of steady-state accretion disks, the matter undergoing accretion in the disk can be characterized by an anisotropic fluid with the energy-momentum tensor \cite{Thorne:1974ve,Page:1974he}
\be
\label{13}
T_{\m\n}=\epsilon_{0}u_{\m}u_{\n}+2u_{(\m}q_{\n)}+t_{\m\n},
\ee
where the terms $\epsilon_{0}$, $q_{\mu}$, and $t_{\mu\nu}$ represent the rest mass density, the energy flow vector, and the stress tensor, respectively. These quantities are defined in the averaged rest-frame of the orbiting particle with the four-velocity $u_{\mu}$. In the rest-frame, $t_{\m\n}$ and $q_{\m}$ are orthogonal to $u_{\mu}$ ($u^{\m}q_{\m}=0$ and $u^{\m}t_{\m\n}$=0). Furthermore, the motion of particles along geodesics ensures that the gravitational pull of the central BH dominates the forces due to radial pressure gradients, and the specific internal energy of the accreting fluid can be ignored compared to its rest energy.

To calculate the flux and  the luminosity emanating from the accretion disc, we assume that the accreting fluid adheres to the conservation of mass, energy, and angular momentum, which take the form, respectively
\be
\1\{\begin{split}
\label{14}
&\dot{M}=-2\pi \sqrt{-\tilde{g}}\Sigma(r) u^{r}=const,\\
&\(\dot{M}\tilde{E}-2\pi\sqrt{-\tilde{g}}\Omega W^{r}_{\phi}\)_{,r}=2\pi\sqrt{-\tilde{g}}F(r)\tilde{E},\\
&\(\dot{M}\tilde{L}-2\pi\sqrt{-\tilde{g}} W^{r}_{\phi}\)_{,r}=2\pi\sqrt{-\tilde{g}}F(r)\tilde{L},
\end{split}\2.
\ee
with
\be
\1\{\begin{split}
\label{15}
&\Sigma(r)=\int^{H}_{-H}\langle\epsilon_{0}\rangle\dif z,\\
&W^{r}_{\phi}=\int^{H}_{-H}\langle t^{r}_{\phi}\rangle\dif z,\\
&\sqrt{-\tilde{g}}=\sqrt{-\det{\tilde{g}_{\m\n}}},
\end{split}\2.
\ee
where $\Sigma(r)$ and $W^{r}_{\phi}$ are the averaged rest mass density and the averaged stress tensor, respectively. $\tilde{g}$ corresponds to the determinant of the near-equatorial metric in cylindrical coordinates and the quantity $\langle t^{r}_{\phi}\rangle$ denotes the average value of the stress tensor during a certain period. Utilizing the energy-angular momentum relation for circular geodesic orbits $\tilde{E}_{,r}=\Omega \tilde{L}_{,r}$, one can eliminate $W^{r}_{\phi}$ from Eq.\eqref{14}. Subsequently, this enables the derivation of the expression for the energy flux.
\be
\label{16}
F(r)=-\frac{\dot{M}}{4 \pi \sqrt{-\tilde{g}}} \frac{\Omega_{, r}}{(\tilde{E}-\Omega \tilde{L})^2} \int_{r_{isco}}^r(\tilde{E}-\Omega \tilde{L}) \tilde{L}_{, r} \dif r.
\ee
Specifically, the above expression is only applicable for cylindrical coordinates. If we want to use it for spherical coordinates, it takes the form \cite{Collodel:2021gxu}
\be
\label{17}
F(r)=-\frac{c^2\dot{M}}{4 \pi \sqrt{-g/g_{\th\th}}} \frac{\Omega_{, r}}{(\tilde{E}-\Omega \tilde{L})^2} \int_{r_{isco}}^r(\tilde{E}-\Omega \tilde{L}) \tilde{L}_{, r} \dif r,
\ee
where we have recovered the dimensions, and $c$ denotes the speed of light. In this scenario, we examine mass accretion propelled by a BH with a total mass of $M=2\times10^6$ $M_{\odot}$ and a corresponding mass accretion rate  $\dot{M}=2\times10^{-6}$ $M_{\odot}yr^{-1}$. In units of the Eddington accretion rate, we have $\dot{M} = 3.36 \times 10^{-4}$ $\dot{M}_{\text{edd}}$, which falls within the range typical for supermassive black holes. For an accreting black hole, the Eddington mass accretion rate is defined as $\dot{M}_{\text{edd}}=\frac{L_{edd}}{\epsilon c^2}$, where $\epsilon$ is the radiative efficiency. The Eddington luminosity is given by
\be
L_{edd}=\frac{4\pi G M m_p c}{\sigma_{th}}=1.26\times10^{38}\(\frac{M}{M_{\odot}}\)ergs^{-1},
\ee
where $m_p$ is the mass of the proton, and $\sigma_{th}$ is the Thomson cross-section.

Figure 3 shows the energy flux $F(r)$ of the disk around a Kerr-Sen BH for different dilaton values $r_2$. We observe that the distribution of the entire energy flux exhibits a trend of initially increasing, reaching a peak, and finally decreasing. The horizontal axis represents the dimensionless coordinate distance $r$, while the vertical axis denotes the order of magnitude of the energy flux. We also conclude that as the parameter $r_2$ increases within the allowed range, the energy flux also increases, surpassing that of the Kerr BH.
\begin{figure}[H]
\centering
\begin{minipage}{0.5\textwidth}
\centering
\includegraphics[scale=0.8,angle=0]{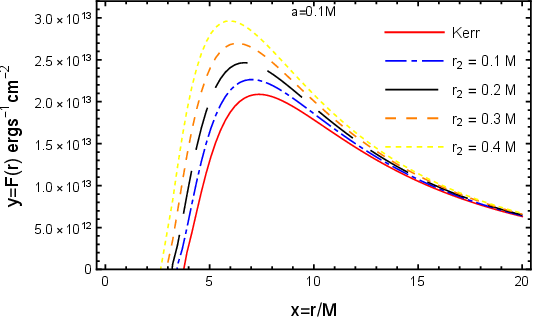}
\end{minipage}%
\begin{minipage}{0.5\textwidth}
\centering
\includegraphics[scale=0.8,angle=0]{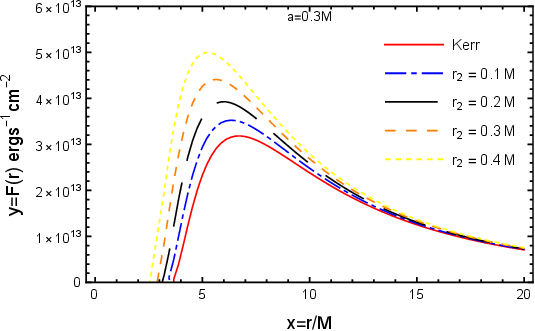}
\end{minipage}
\begin{minipage}{0.5\textwidth}
\centering
\includegraphics[scale=0.8,angle=0]{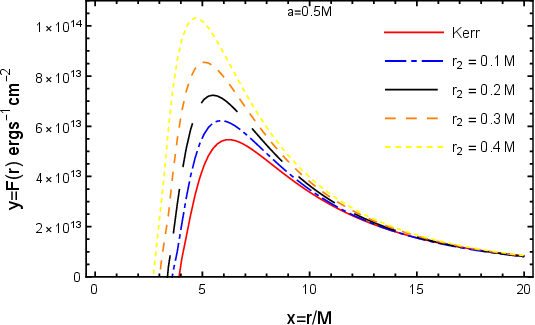}
\end{minipage}%
\begin{minipage}{0.5\textwidth}
\centering
\includegraphics[scale=0.8,angle=0]{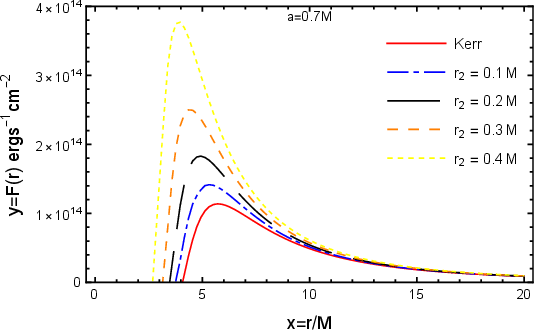}
\end{minipage}
\caption{\label{Fig.3} The energy flux $F(r)$ of accretion disk around a Kerr-Sen BH for different values of the dilaton parameter $r_2$ with $M_{\odot}= 1.989 \times10^{33}$ g, $\frac{a}{M}=0.1$ (top-left), $\frac{a}{M}=0.3$ (top-right), $\frac{a}{M}=0.5$ (bottom-left), and $\frac{a}{M}=0.7$ (bottom-right). The red line represents the radiation scenario for the Kerr BH.   }
\end{figure}

Within the framework of the Novikov-Thorne model, the accreted matter is assumed to be in thermodynamic equilibrium. This implies that the radiation emitted by the disk can be treated as perfect black body radiation. The radiation temperature $T(r)$ of the disk is related to the energy flux $F(r)$ through the Stefan-Boltzmann law, expressed as
 \be
F(r)=\sigma_{sb}T^4(r),
\ee
where $\sigma_{sb}$ denotes the Stefan-Boltzmann constant. This suggests that the variation of $T(r)$ with respect to
$r$ is similar to the dependence of the energy flux $F(r)$ on $r$. In Fig. 4, we plot the radiation temperature $T(r)$ of the thin disk around a Kerr-Sen BH for fixed dilaton parameter $r_2$ and rotation parameter $a$. Similar to the result for the energy flux, we find that as the parameter $r_2$ increases, the radiation temperature also increases, reaching its maximum value. Subsequently, the temperature decreases as the the distance $r$ increase.
\begin{figure}[H]
\centering
\begin{minipage}{0.5\textwidth}
\centering
\includegraphics[scale=0.8,angle=0]{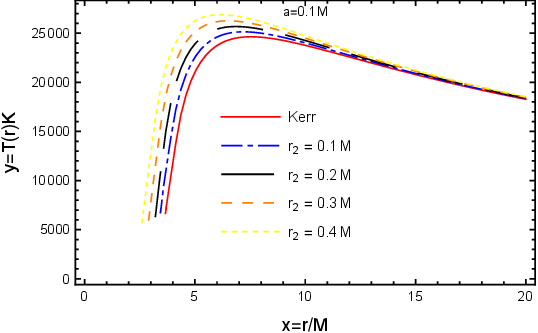}
\end{minipage}%
\begin{minipage}{0.5\textwidth}
\centering
\includegraphics[scale=0.8,angle=0]{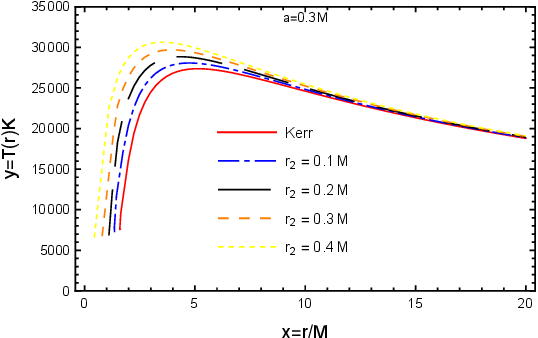}
\end{minipage}
\begin{minipage}{0.5\textwidth}
\centering
\includegraphics[scale=0.8,angle=0]{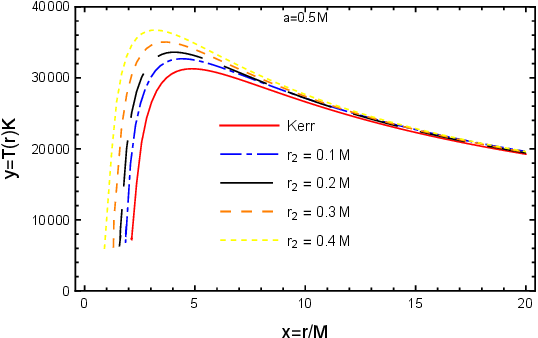}
\end{minipage}%
\begin{minipage}{0.5\textwidth}
\centering
\includegraphics[scale=0.8,angle=0]{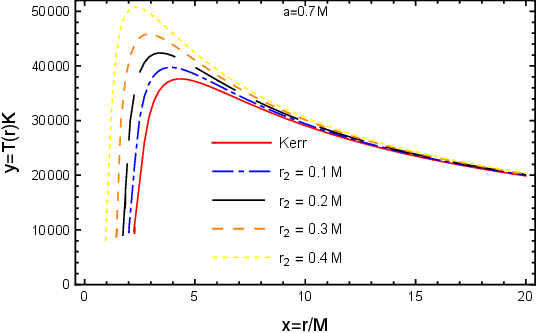}
\end{minipage}
\caption{\label{Fig.4} The temperature $T(r)$ of accretion disk around a Kerr-Sen BH for different values of the dilaton parameter $r_2$ with $\sigma_{sb}=5.67\times10^{-5}$ erg$s^{-1}$c$m^{-2}K^{-4}$, $\frac{a}{M}=0.1$ (top-left), $\frac{a}{M}=0.3$ (top-right), $\frac{a}{M}=0.5$ (bottom-left), and $\frac{a}{M}=0.7$ (bottom-right), respectively.  }
\end{figure}

The observed luminosity and the efficiency of matter accretion stand as crucial measurement parameters in the accretion process. The luminosity $L(\n)$ is associated with the alteration in frequency of a photon as it travels from the emitting source to the observer \cite{Torres:2002td,Banerjee:2020qmi}. The maximum efficiency $\epsilon$ is determined by the specific binding energy at the marginally stable orbit $r_{isco}$ \cite{Page:1974he, Novikov:1973kta,Banerjee_2021}. These two physical quantities are
\be
\1\{\begin{split}
\label{18}
& L(\nu)=4 \pi d^2 I(\nu)=\frac{8 \pi h \cos \gamma}{c^2} \int_{r_i}^{r_f} \int_0^{2 \pi} \frac{\nu_e^3 r \dif r \dif \phi}{e^{h \nu_e / k T(r)}-1},  \\
& \epsilon=1-\tilde{E}_{isco}(r_{isco}).
\end{split}\2.
\ee
The emitted frequency is denoted by $\n_{e}=\n(1+z)$, and the redshift factor $g$ can be defined as
\be
\label{19}
g\equiv1+z=\frac{1+\Omega r \sin \phi \sin \gamma}{\sqrt{-g_{tt}-2 \Omega g_{t\phi}-\Omega^2 g_{\phi\phi}}}.
\ee
where $\tilde{E}_{isco}$ is the specific energy of the test particle at $r_{isco}$, $h$ is Planck constant, and $k$ is the Boltzmann constant. The $\gamma$ and $d$ represents the inclination angle of the accretion disk and the distance from the observer to the disk center, respectively. $L(\n)$ is the thermal energy flux radiated by the disk. Notably, we ignore light bending, which may be an acceptable approximation only for low values of the inclination angle $\gamma$. To determine the luminosity $L(\n)$ of the disk, we consider $r_{i}=r_{isco}$ and $r_{f}=\infty$ since the flux over the disk surface decreases as $r\rightarrow\infty$ in the rotating Kerr-Sen BH.
\begin{figure}[H]
\centering
\begin{minipage}{0.5\textwidth}
\centering
\includegraphics[scale=0.8,angle=0]{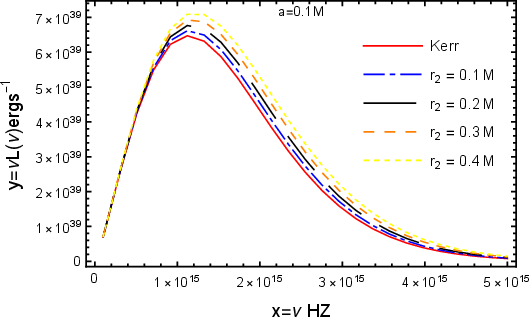}
\end{minipage}%
\begin{minipage}{0.5\textwidth}
\centering
\includegraphics[scale=0.8,angle=0]{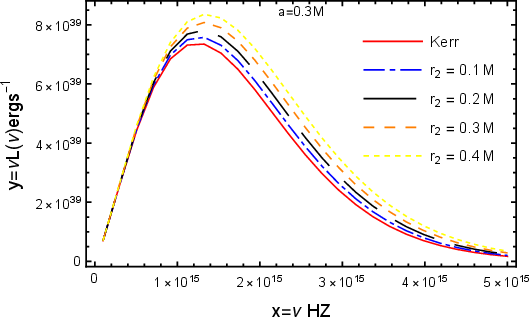}
\end{minipage}
\begin{minipage}{0.5\textwidth}
\centering
\includegraphics[scale=0.8,angle=0]{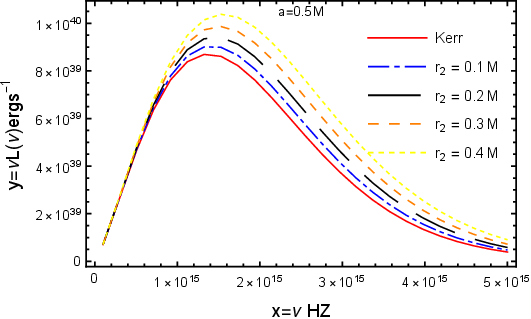}
\end{minipage}%
\begin{minipage}{0.5\textwidth}
\centering
\includegraphics[scale=0.8,angle=0]{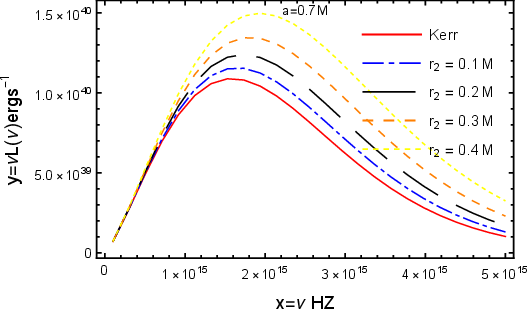}
\end{minipage}

\caption{\label{Fig.5}The emission spectrum $\n L(\n)$ of accretion disk around Kerr-Sen BH for different $r_2$ with $\frac{a}{M}=0.1$ (top-left), $\frac{a}{M}=0.3$ (top-right), $\frac{a}{M}=0.5$ (bottom-left), and $\frac{a}{M}=0.7$ (bottom-right). The inclination angle is $\gamma=0^{\circ}$.  }
\end{figure}

The variations of spectral energy distribution are illustrated in Fig. 5. The two figures illustrate the variations in $\n L(\n)$ with $\frac{a}{M}=0.1,0.3,0.5,0.7$, and different dilaton parameter $r_2$. Initially, the entire spectrum exhibits a blackbody distribution with the peak occurring at higher frequencies. However, we can't distinguish different BHs at lower frequencies. Additionally, as the parameter $r_2$ increases, a noticeable rightward shift in the peak is observed, which indicates greater emitted energy compared to the Kerr BH. This conclusion is consistent with the Ref. \cite{Banerjee:2020qmi}, though the luminosity formula and the consideration of model parameters differ.

In our previous work, we have provided conclusion on the radiation efficiency \cite{Feng:2024mey}. It shows that the radiative efficiency of the Kerr-Sen BH increases with the growth of the parameter $r_2$ (or the parameter $a$). In addition, the efficiency of the Kerr-Sen BH surpasses that of the Kerr BH with fixed parameter. This observation contributes to a deeper understanding of these two BHs through empirical observations.

\section{Black hole shadow in plasma environment}
In this section, we will explore the shadow of the Kerr-Sen BH within a homogeneous plasma environment. The study of radiation propagation through an isotropic and dispersive medium in GR was initially conducted by Bicak and Hadrava in 1975 \cite{Kichenassamy:1985zz}. Additionally, the shadow of various BHs in the presence of plasma has been explored in Refs.\cite{ Bisnovatyi-Kogan:2015dxa,Er:2013efa,Sahoo:2023czj,Wei:2013kza,Jana:2023sil,Dastan:2016bfy,Lan:2018lyj,Xavier:2020egv,Wu:2022ydk,Uniyal:2022vdu,Cao:2023ppv,Badia:2022phg,Jusufi:2020dhz,Feng:2019zzn,Vagnozzi:2022moj,Kumar:2024vdh,Kumar:2023wfp, Kim:2020bhg,Ovgun:2018tua,Sui:2023rfh,Atamurotov:2022nim,Xavier:2020egv,Mirzaev:2023oud,Hu:2024cbn,Ghosh:2022kit,10.1093/pasj/57.2.273,Hou2022MultilevelIA,Nedkova2023,PhysRevD.97.064021,PhysRevD.109.024060}. Here, we examine the effect of the plasma parameter on the image of shadow of the Kerr-Sen BH.
\subsection{Hamilton-Jacobi equation for null geodesics in plasma }
We consider a straightforward scenario involving a cold (pressureless) and nonmagnetized plasma, where the behavior of light is governed by the Hamiltonian \cite{Perlick:2017fio,Jana:2023sil,Dastan:2016bfy,Kumar:2023wfp,Kumar:2024vdh}
\be
\label{20}
H(x,p)=\frac{1}{2}\(g^{\m\n}p_{\m}p_{\n}+\omega^2_{p}(x) \),
\ee
where $\omega_{p}$ represents the plasma electron frequency given by
\be
\label{21}
\omega^2_{p}(x)\equiv\frac{4\pi e^2}{m_{e}}N(x),
\ee
$m_e$ and $e$ represents the mass and charge of the electron, respectively, while $N(x)$ describes the distribution of electron density. Here, $x$ denotes the spacetime coordinates $\(t,r,\th,\phi\)$ and $p$ refers to the momentum coordinates $\(p_{t},p_{r},p_{\th},p_{\phi}\)$. Additionally, the plasma frequency $\omega_{p}$ and the photon frequency $\omega$ are related by refractive index $n$, which can be defined as
\be
\label{22}
n^2\equiv 1-\frac{\omega^2_{p}(x)}{\omega^2(x)}.
\ee

Taking into account the gravitational redshift effect on the constant of motion $p_{t}\equiv-\omega_{0}$, the frequency of the observed redshift is given by $\omega(x)=\frac{\omega_{0}}{\sqrt{-g_{00}}}$. Since the refractive index
$n>0$,  the condition for generalized rotating metric can be given by \cite{Kumar:2023wfp,Kumar:2024vdh}
\be
\label{23}
\omega^2_{0}>-g_{tt} \omega^2_{p}(x).
\ee

We will explore the homogeneous plasma distribution described by the equation $\omega^2_{p}=k \omega^2_{0}$ spanning the region between the observer and the horizon. Here, $k$ represents the homogeneous plasma parameter, which must fall within the interval ($0<k<1$) to adhere to Eq. \eqref{23}. Subsequently, let's consider the Hamilton-Jacobi equation for the null geodesics as
\be
\label{24}
H\(x,\frac{\p S}{\p x^{\m}}\)=\frac{1}{2}g^{\m\n}\frac{\p S}{\p x^{\m}}\frac{\p S}{\p x^{\n}}+\frac{1}{2}\omega^2_{p}(x)=0,
\ee
with the separation ansatz
\be
\label{25}
S=-\omega_{0}t+L_{z}\phi+S_{r}(r)+S_{\th}(\th),
\ee
where $L_{z}$ represents conserved quantity of photons along the $\phi$ direction. Given that our study focuses on the variation of coordinates $r$ and $\th$ within a plasma environment, separation of the above equation is feasible only if we adopt a general expression for the plasma frequency in the Kerr-Sen BH, as follows \cite{PhysRevD.95.104003,Junior:2020lya}
\be
\label{26}
\omega^2_{p}=\frac{f_{r}(r)+f_{\th}(\th)}{\tilde{\Sigma}}.
\ee
In the uniform plasma system, $f_{r}(r)$ and $f_{\th}(\th)$ can each be assigned the values $k\omega^2_{0}r(r+r_2)$ and $k\omega^2_{0}a^2\cos^2{\th}$. Now, using Eq. \eqref{24}, Eq. \eqref{25} and Eq. \eqref{26}, we get the following expressions
\be
\1\{\begin{split}
\label{27}
&\tilde{\Delta}^2\(\frac{\dif S}{\dif r}\)^2=-K\tilde{\Delta}-a^2\omega^2_{0}\tilde{\Delta}-\tilde{\Delta}L^2_z-k\omega^2_{0}\tilde{\Delta}[r(r+r_2)]+a^2L^2_z+\omega^2_{0}\[r(r+r_2)+a^2\]^2,  \\
&\(\frac{\dif S}{\dif \th}\)^2=K-k\omega^2_0a^2\cos^2{\th}-L^2_{z}\cot^2{\th}+a^2\omega^2_{0}\cos^2{\th},
\end{split}\2.
\ee
where $K$ is Carter's constant and $\frac{\dif S}{\dif x^{\m}}=p_{\m}$. By employing the Hamiltonian equations, we can simplify the equations of motion to
\be
\1\{\begin{split}
\label{28}
&\frac{\dif t}{\dif\lambda}=\frac{\omega_{0}}{\tilde{\Delta}\tilde{\Sigma}}\[(r(r+r_2)+a^2)^2-\tilde{\Delta}a^2\sin{\th}^2-2Mra \xi   \], \\
&\frac{\dif \phi}{\dif\lambda}=\frac{\omega_{0}}{\tilde{\Delta}\tilde{\Sigma}}\( 2Mra+\frac{\tilde{\Sigma}-2Mr}{\sin{\th}^2}\xi   \), \\
&\frac{\dif r}{\dif\lambda}=\pm\frac{1}{\tilde{\Sigma}}\sqrt{R(r)},\\
&\frac{\dif \th}{\dif\lambda}=\pm\frac{1}{\tilde{\Sigma}}\sqrt{\Theta{(\th)}},
\end{split}\2.
\ee
with
\be
\1\{\begin{split}
\label{29}
&R(r)\equiv\omega^2_{0}\[a^2\xi^2+(r(r+r_2)+a^2)^2-4Mra\xi-\tilde{\Delta}(\eta+a^2+\xi^2+kr(r+r_2)) \],\\
&\Theta{(\th)}\equiv\omega^2_{0}\(\eta+a^2\cos^2{\th}-\xi^2\cot^2{\th}-ka^2\cos^2{\th}  \),
\end{split}\2.
\ee
where $\lambda$ is affine parameter and two impact parameters $ \xi \equiv \frac{L_z}{\omega_{0}}$, $\eta = \frac{K}{\omega^2_{0}}$ characterize the system \cite{Sahoo:2023czj}. The parameter $\xi$ denotes the distance from the axis of rotation, while $\eta$ signifies the distance from the equatorial plane. It is worth noting that when the $k=0$, the equation reduces to the photon trajectories of the vacuum case.

Since the shadow is created by the last photon rings, which are inherently unstable, null rays must meet the following conditions to maintain unstable spherical orbits \cite{PhysRevD.95.104003},
\be
\label{30}
\frac{\dif^2R(r)}{\dif r^2}>0 ,~~~~ \Theta(\th)\geqslant0.
\ee
The first condition is necessary for these orbits to be unstable, while the second condition ensures the existence of spherical orbits around the Kerr-Sen BH. A similar analysis for axially symmetric stationary spacetime can be found in \cite{Bezdekova:2022gib}.
\subsection{The shadow of Kerr-Sen black hole in plasma spacetime}
In this subsection, we will explore the effects of the dilaton parameter $r_2$ and the homogeneous plasma parameters $k$ on the BH's shadow. To streamline our analysis, we derive the general expression for the impact parameters corresponding to the shadow boundary. We analyze the radial motion described by Eq. \eqref{28} to determine the boundary of shadow of Kerr-Sen BH, which is defined by the circular photon orbit satisfying the following conditions \cite{Frolov:1998wf,Teo:2003ltt}
\be
\label{31}
R(r_c)=0,~~~~\frac{\dif R(r_c)}{\dif r}=0,
\ee
where $r_c$ denotes the critical photon orbit. By solving these conditions, we obtain these expressions for the impact parameters as follows
\be
\1\{\begin{split}
\label{32}
&\xi=\frac{A_1+2  M (a-r_c) (a+r_c)}{a \left(2 M-2 r_c-r_2\right)},\\
&\eta=\frac{r^2_c\[4M(A_1+A_2+\frac{1}{2}A_3)+ \left(2 r_c+r_2\right)A_4\]}{a^2 (2 M-2 r_c-r_2)^2},
\end{split}\2.
\ee
where
\be
\1\{\begin{split}
\label{33}
&A_1\equiv\left[a^2+r_c \left(r_c-2 M+r_2\right)\right] \sqrt{\left(2 r_c+r_2\right) \left[k \left(2 M-2 r_c-r_2\right)+2 r_c+r_2\right]},\\
&A_2\equiv M\[(8 k-5) r^2+2 (5 k-2) r_2 r+(3 k-1) r_2^2\],\\
&A_3\equiv a^2\[2 k (M-r_c)-(k-2) r_2+4 r_c\], \\
&A_4\equiv -8 k M^3-2 M \left(r+r_2\right) \left[(5 k-4) r+(3 k-2) r_2\right]+(k-1) \left(r+r_2\right){}^2 \left(2 r+r_2\right).
\end{split}\2.
\ee
As previously discussed, these critical orbits are essential in defining the last photon rings. Consequently, $\xi$ and $\eta$ can entirely delineate the shadow's boundary. Nonetheless, to visualize the shadow in the observer's sky, we utilize the following celestial coordinates \cite{Vazquez:2003zm}
\be
\1\{\begin{split}
\label{34}
&\al=\lim_{r_0\rightarrow\infty}\(r^2_0\sin{\th}\frac{\dif\phi}{\dif r}\Big|_{\th\rightarrow\th_0}\),\\
&\beta=\lim_{r_0\rightarrow\infty}\(r^2_0\frac{\dif\th}{\dif r} \Big|_{\th\rightarrow\th_0}\),
\end{split}\2.
\ee
where $r_0$ represents the distance from the distant observer to the BH, and $\th_0$ represents the inclination angle between the observer's line of sight and the BH's spin axis. The parameters $\al$ and $\beta$ give the shadow's profile. Using the geodesic equation \eqref{28} and applying the condition for unstable circular orbit, we can simplify the celestial coordinates as follows
\be
\1\{\begin{split}
\label{35}
&\al=-\frac{\xi}{\sin{\th_0}\sqrt{1-k}},\\
&\beta=\pm\frac{\sqrt{\eta-\xi^2\cot{\th^2_0}+(1-k)a^2\cos^2{\th_0}}}{\sqrt{1-k}}.
\end{split}\2.
\ee
It is crucial to recognize that these expressions do not apply in a vacuum distribution ($k=0$). In particular, for a uniform distribution, the plasma density is pivotal in determining the shape of a BH's shadow. The panel (left) of Fig. 6 illustrates the variation of the BH's shadow distribution in a uniform plasma environment with respect to the parameter $k$, while the panel (right) depicts the variation of the shadow distribution with the parameter $r_2$. We observe that as the parameter $k$ increases, the shadow gradually enlarges, while with the increase of the parameter $r_2$, the shadow decreases. These effects are also likely to be easily observed by low-resolution radio telescopes.
\begin{figure}[H]
\centering
\begin{minipage}{0.5\textwidth}
\centering
\includegraphics[scale=0.8,angle=0]{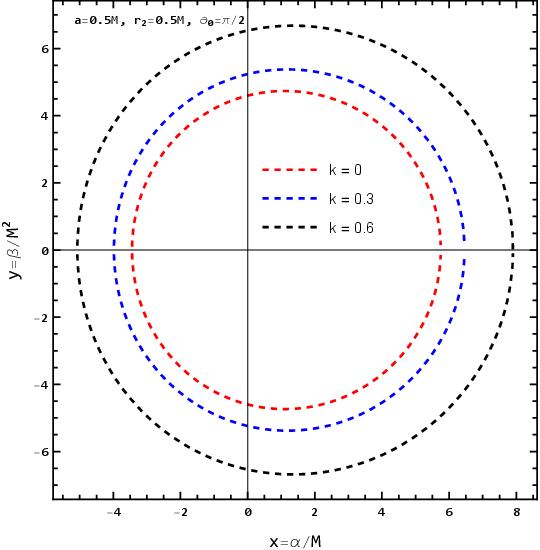}
\end{minipage}%
\begin{minipage}{0.5\textwidth}
\centering
\includegraphics[scale=0.8,angle=0]{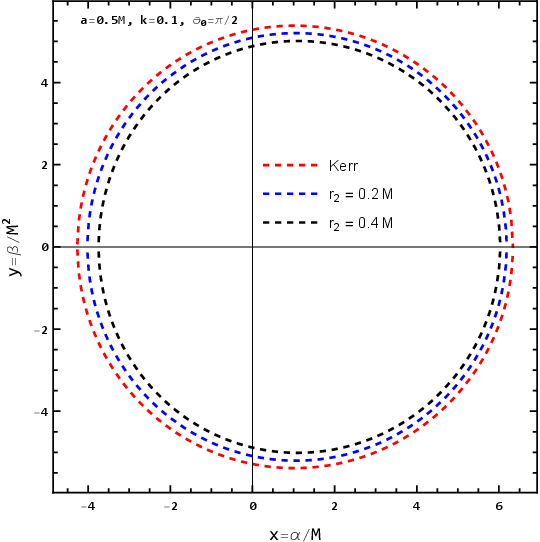}
\end{minipage}
\caption{\label{Fig.6}
The figures depict the images of shadow with $\frac{r_2}{M}=0.5$, $\th_0 = 90^{\circ}$, $\frac{a}{M}= 0.5$, and some values of $k$ for the left panel, with fixed $\th_0$, $k$, $a$, and different values of $r_2$ for the right panel.}
\end{figure}
\section{Constraints on parameters with the EHT observations of M87* and Sgr A*}
The EHT observations of  M87* and Sgr A* provide constraints on theoretical models of BHs. By capturing the shadow and surrounding emission of these supermassive BHs, the EHT data allows researchers to constrain parameters. These observations offer unprecedented insights into the nature of BHs and their immediate environments. In this section, we will utilize observational data from the EHT concerning the supermassive BH at the M87* and Sgr A* to constrain Kerr-Sen and Kerr-Newman BHs.

According to Refs. \cite{EventHorizonTelescope:2019dse,EventHorizonTelescope:2019ths,EventHorizonTelescope:2019ggy}, the angular diameter of the shadow  of M87* is $\Delta\th_{M87^*}=42\pm3$ $\mu as$, the distance from Earth to the BH is $D=16.8$ Mpc, and the mass of M87* is $M=(6.5\pm0.9)\times10^{9}$ $M_{\odot}$. Conversely, the data for Sgr A* is sourced from the recent literatures \cite{EventHorizonTelescope:2022wkp,Wang:2022ivi}, which report the angular diameter of Sgr A* is $\Delta\th_{Sgr A^*}=48.7\pm7$ $\m as$, the distance is $D=8277\pm33$ pc, and the mass is $M =(4.3\pm0.013)\times10^{6}$ $M_{\odot}$. Given these data, Refs. \cite{Kumar:2023wfp,Kumar:2024vdh,Uniyal:2022vdu} estimate the diameter of the BH shadow $d_{sh}$, which can be expressed as
\be
\1\{\begin{split}
\label{36}
&d^{M87^*}_{sh}=2\times D\times\tan{\frac{\Delta\th_{M87^*}}{2}}\approx(11\pm1.5) M,\\
&d^{Sgr A^*}_{sh}=2\times D\times\tan{\frac{\Delta\th_{Sgr A^*}}{2}}\approx(9.5\pm1.4) M.
\end{split}\2.
\ee
Subsequently, to match the theoretical shadow diameter with observational data, we need to consider the radius $R_{sh}$ of shadow of Kerr-Sen BH. In this analysis, we consider the case of an inclination angle of $\th_{0}=90^\circ$. According to Refs. \cite{Feng:2019zzn,Jusufi:2020dhz,Wu:2021pgf}, the radius of shadow of Kerr-Sen BH can be defined as
\be
\1\{\begin{split}
\label{37}
&R_{sh}=\frac{1}{2}\[\al(r^{+}_{sp})-\al(r^{-}_{sp}) \],\\
&\beta(r^{\pm}_{sp})=0,
\end{split}\2.
\ee
where $r^{\pm}_{sp}$ ($r^+_{sp}>r^-_{sp}$) denotes the unstable photon orbits. It is noteworthy that photons at $\al(r^{+}_{sp})$/$\al(r^{-}_{sp})$ typically originate from orbits rotating in the same/opposite directions of the BH. Hence, the theoretical shadow diameter can be obtained via $d^{theo}_{sh}=2R_{sh}$. Furthermore, by comparing the theoretical value with the observational data, we can constrain the range of the model parameters.

Figure 7 illustrates the variation of the diameter of shadow as a function of the parameters $k$ and $r_2$ for Kerr-Sen BH with constraints from M87* and Sgr A*. The region outlined by black dashed lines represents the $1\sigma$ confidence interval, while the area enclosed by blue dashed lines indicates the $2\sigma$ confidence interval. The top two panels illustrate the variations in the diameter of shadow of Kerr-Sen BH with the dilaton parameter $r_2$ in a vacuum environment ($k=0$) for fixed parameter $a$. The bottom two panels depict the change in the diameter of shadow of Kerr-Sen BH with the homogeneous plasma parameter $k$. Similarly, Figure 8 illustrates the constraints on the Kerr-Newman BH, where we adopt the results derived in Ref. \cite{Xavier:2020egv}. The intersections of the dashed and solid lines indicate the upper limit of the parameter, which are provided in Tab. 2 and Tab. 3, respectively. It can be observed that in the vacuum case, the upper bound of the dilaton parameter $r_2$ decreases as the spin parameter $a$ increases. Furthermore, for M87* data with $\frac{r_2}{M}=0.1$, the homogeneous plasma parameter $k$ decreases with increasing $a$ within the $1\sigma$ interval and is allowed within the $2\sigma$ interval. For Sgr A* data, the parameter $k$ is permitted within both the $1\sigma$ and  $2\sigma$ intervals. For the case of the Kerr-Newman BH, the data from M87* and Sgr A* do not provide any constraints on the parameters within the 2$\sigma$ range, while constraints are given within the 1$\sigma$ range. Furthermore, within the 1$\sigma$ interval for a fixed $a$ parameter, the constrained range of the dilaton charge in the Kerr-Sen solution is slightly higher than the electric charge in the Kerr-Newman BH, with almost no impact on the constraint for $k$. The differences in the constrained ranges between two BHs are minimal, which may pose challenges in distinguishing them observationally.

\begin{figure}[H]
\centering
\begin{minipage}{0.48\textwidth}
\centering
\includegraphics[scale=0.8,angle=0]{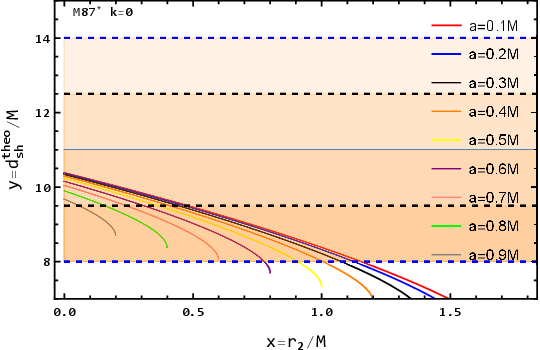}
\end{minipage}%
\begin{minipage}{0.495\textwidth}
\centering
\includegraphics[scale=0.8,angle=0]{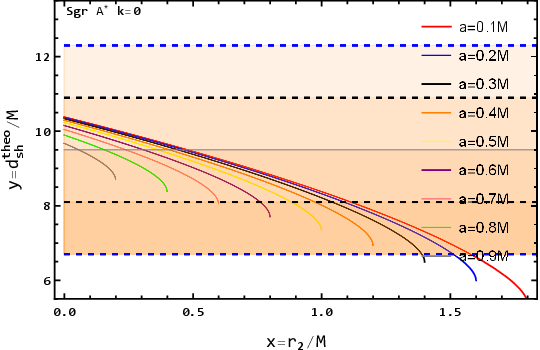}
\end{minipage}
\centering
\begin{minipage}{0.48\textwidth}
\includegraphics[scale=0.81,angle=0]{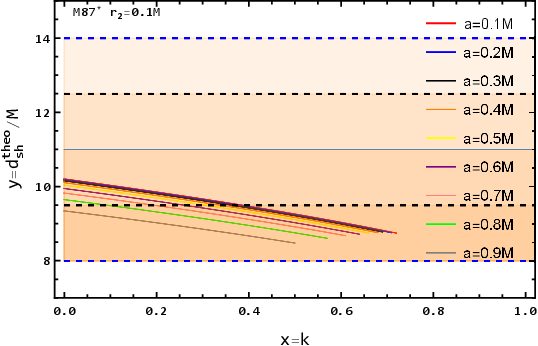}
\end{minipage}%
\begin{minipage}{0.47\textwidth}
\centering
\includegraphics[scale=0.81,angle=0]{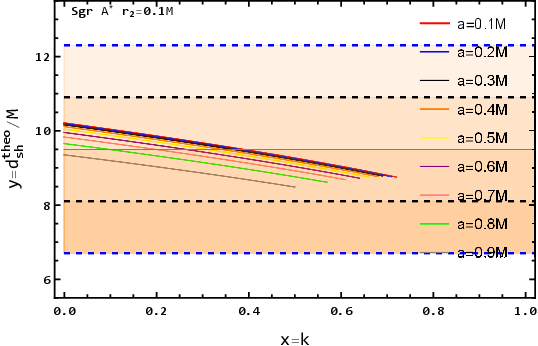}
\end{minipage}
\caption{\label{Fig.7}The top two plots illustrate the constraints on the dilaton parameter $r_2$ within $1\sigma$ (bounded by the two black dashed lines) and $2\sigma$ (bounded by the two blue dashed lines) confidence intervals for different values of $a$ in a vacuum environment ($k=0$). The bottom two plots depict the constraints on the homogeneous plasma parameter $k$ with fixed parameter $\frac{r_2}{M}=0.1$. }
\end{figure}

\begin{figure}[H]
\centering
\begin{minipage}{0.48\textwidth}
\centering
\includegraphics[scale=0.8,angle=0]{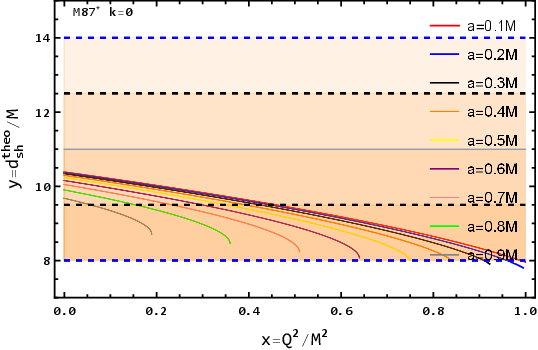}
\end{minipage}%
\begin{minipage}{0.495\textwidth}
\centering
\includegraphics[scale=0.8,angle=0]{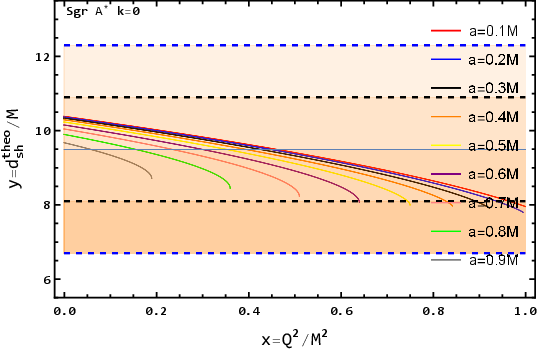}
\end{minipage}
\centering
\begin{minipage}{0.48\textwidth}
\includegraphics[scale=0.81,angle=0]{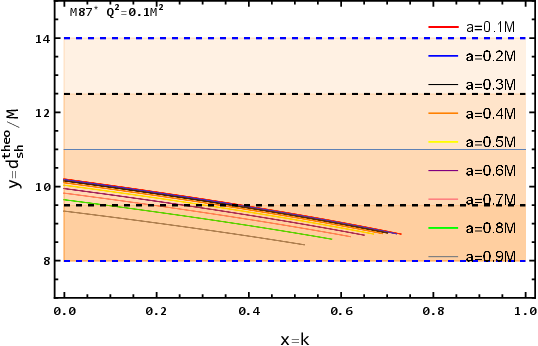}
\end{minipage}%
\begin{minipage}{0.47\textwidth}
\centering
\includegraphics[scale=0.81,angle=0]{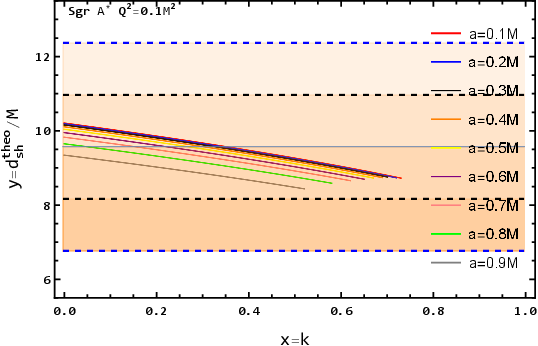}
\end{minipage}
\caption{\label{Fig.8}The top two panels  represent the constraints on the charge $Q$ of the Kerr-Newman black hole, while the bottom two panels illustrate the constraint on the plasma parameter $k$.    }
\end{figure}

\begin{table}[H]
    \centering
     \caption{\label{Tab.1}The constraints on the parameters from M87* data. The dash indicates the absence of a corresponding parameter value.    }
   \scalebox{0.8}{
    \begin{tabular}{|llccc|ccccc|}
        \hline
        \hline
       (Kerr-Sen) M87*&   &  1$\sigma$&     & 2$\sigma$                                 &  (Kerr-Sen)    M87*      &      &1$\sigma$  & &    2$\sigma$       \\
        \hline
        $\frac{a}{M}$  & $k$         &   Upper bound $\(\frac{r_2}{M}\)$           &&  Upper bound  $\(\frac{r_2}{M}\)$         & $\frac{a}{M}$  & $\frac{r_2}{M}$             &    Upper bound $\(k\)$            && Upper bound $\(k\)$ \\
           \hline
         0.1&  0                           &   0.480718                                  &&   1.153510              &      0.1  &0.1           & 0.387790                              &             & -               \\
         0.2&  0                           &   0.466644                                  &&   1.125200              &      0.2  &0.1           & 0.378870                              &             & -                \\
         0.3&  0                           &   0.442937                                  &&   1.076700              &      0.3  &0.1           & 0.363370                              &             & -                 \\
         0.4&  0                           &   0.409200                                  &&   1.005730              &      0.4  &0.1            &0.340190                              &             & -                 \\
         0.5&  0                           &   0.364832                                  &&   0.908329              &      0.5  &0.1            &0.307390                              &             & -            \\

         0.6&  0                           &   0.308967                                 &&   0.777482              &      0.6  &0.1             &0.261520                             &             & -            \\
         0.7&  0                           &   0.240382                                 &&   -                           &      0.7  &0.1            &0.195910                              &             & -            \\
         0.8&  0                           &   0.157328                                 &&  -                            &      0.8  &0.1            &0.094890                              &             & -            \\
         0.9&  0                           &   0.057232                                 &&   -                           &      0.9  &0.1            &-                                        &             & -            \\
    \hline
   (Kerr-Newman)   M87*&   &  1$\sigma$&     & 2$\sigma$                                 &(Kerr-Newman)      M87*      &      &1$\sigma$  & &    2$\sigma$       \\
        \hline
   $\frac{a}{M}$  & $k$         &   Upper bound $\(\frac{Q^2}{M^2}\)$           &&  Upper bound  $\(\frac{Q^2}{M^2}\)$         & $\frac{a}{M}$  & $\frac{Q^2}{M^2}$             &    Upper bound $\(k\)$            && Upper bound $\(k\)$ \\
           \hline
         0.1&  0                           &   0.458875                                  &&   -                         &      0.1  &0.1           & 0.387144                              &             & -               \\
         0.2&  0                           &   0.445750                                  &&   -                         &      0.2  &0.1           & 0.378205                              &             & -                \\
         0.3&  0                           &   0.423630                                  &&   -                         &      0.3  &0.1           &  0.362670                             &             & -                 \\
         0.4&  0                           &   0.392120                                  &&   -                         &      0.4  &0.1            & 0.339420                            &             & -                 \\
         0.5&  0                           &   0.350603                                  &&   -                         &      0.5  &0.1            & 0.306510                           &             & -            \\

         0.6&  0                           &  0.298142                                   &&  -                          &      0.6  &0.1             &0.260480                            &             & -            \\
         0.7&  0                           &  0.233336                                   &&   -                         &      0.7  &0.1            &  0.194570                               &             & -            \\
         0.8&  0                           &  0.154030                                   &&   -                         &      0.8  &0.1            &  0.092920                                 &             & -            \\
         0.9&  0                           &  0.056743                                   &&   -                         &      0.9  &0.1            &-                                        &             & -            \\
      \hline

    \end{tabular}}
   \end{table}

\begin{table}[H]
    \centering
\caption{\label{Tab.2}The constraints on the parameters from Sgr A* data.                   }
       \scalebox{0.78}{
    \begin{tabular}{|llccc|ccccc|}
        \hline
        \hline
      (Kerr-Sen)  Sgr A*&   &  1$\sigma$&     & 2$\sigma$             &  (Kerr-Sen)  Sgr A*    &      &1$\sigma$  & &    2$\sigma$       \\
        \hline
        $\frac{a}{M}$  & $k$  &    Upper bound $\(\frac{r_2}{M}\)$           &&  Upper bound  $\(\frac{r_2}{M}\)$         & $\frac{a}{M}$  & $\frac{r_2}{M}$             &    Upper bound  $\(k\)$            &&  Upper bound $\(k\)$ \\
           \hline
         0.1&  0                           & 1.114420                                 &&   1.579320                     &      0.1  &0.1           & -                                && -             \\
         0.2&  0                           & 1.087590                                 &&   1.512470                     &      0.2  &0.1           & -                                && -              \\
         0.3&  0                           & 1.041720                                 &&   1.389720                     &      0.3  &0.1           & -                                && -               \\
         0.4&  0                           & 0.974817                                 &&   -                                  &      0.4  &0.1           & -                               && -                 \\
         0.5&  0                           & 0.883507                                 &&  -                                   &      0.5  &0.1           & -                               &&-                 \\

         0.6&  0                           & 0.761949                                &&  -                                   &      0.6  &0.1           & -                               &&-                 \\
         0.7&  0                           & 0.599166                                &&  -                                   &      0.7  &0.1           & -                               &&-                 \\
         0.8&  0                           & -                                             &&  -                                   &      0.8  &0.1           & -                               &&-                 \\
         0.9&  0                           & -                                              &&  -                                   &      0.9  &0.1           & -                               &&-                 \\
  \hline
   (Kerr-Newman)   Sgr A*&   &  1$\sigma$&     & 2$\sigma$                                 &(Kerr-Newman)      Sgr A*      &      &1$\sigma$  & &    2$\sigma$       \\
        \hline
        $\frac{a}{M}$  & $k$         &   Upper bound $\(\frac{Q^2}{M^2}\)$           &&  Upper bound  $\(\frac{Q^2}{M^2}\)$         & $\frac{a}{M}$  & $\frac{Q^2}{M^2}$             &    Upper bound $\(k\)$            && Upper bound $\(k\)$ \\
           \hline
         0.1&  0                           &   0.965120                                  && -                &      0.1  &0.1           & -                              &             & -               \\
         0.2&  0                           &   0.937782                                 && -                &      0.2  &0.1           & -                              &             & -                \\
         0.3&  0                           &   0.892148                                              &&   -              &      0.3  &0.1           & -                              &             & -                 \\
         0.4&  0                           &   0.828006                                        &&  -               &      0.4  &0.1            &-                              &             & -                 \\
         0.5&  0                           &   0.744660                                         &&  -               &      0.5  &0.1            &-                              &             & -            \\

         0.6&  0                           &  0.639683                                  &&   -               &      0.6  &0.1             &-                             &             & -            \\
         0.7&  0                           &   -                               &&   -               &      0.7  &0.1            &-                              &             & -            \\
         0.8&  0                           &    -                              &&  -                &      0.8  &0.1            &-                              &             & -            \\
         0.9&  0                           &     -                              &&   -               &      0.9  &0.1            &-                               &             & -            \\
  \hline
    \end{tabular}}
   \end{table}

\section{Conclusion and discussion}

In our investigation of rotating Kerr-Sen BH within the EMDA framework, we focused on exploring the characteristics of thin relativistic accretion disks using the Novikov-Thorne model. Initially, we derived key physical quantities, including the effective potential $V_{\text{eff}}$, the specific angular momentum $\tilde{L}$, the specific energy $\tilde{E}$, the angular velocity $\Omega$, and the ISCO radius for test particles in circular orbits around the BH. Since the ISCO radius lacked an analytical solution, we employed numerical methods to calculate it for various spin $a$ and dilaton parameters $r_2$. Our analysis revealed that, for a given $a$ value, the ISCO radius decreases as the dilaton parameter increases. However, the specific angular momentum $\tilde{L}$ and the specific energy $\tilde{E}$ of detected particle undergo changes with the dilaton parameter $r_2$: when the parameter exceeds a critical value, the function transits from a decreasing to an increasing trend. Subsequently, the particle's angular velocity $\Omega$ exhibits a continuous decrease as $r_2$ increases.

Additionally, we performed numerical calculations to determine the energy flux $F(r)$, the temperature $T(r)$, and the luminosity distribution $L(\nu)$ emitted by the accretion disk, and created diagrams to visualize these quantities. Our results highlight the influence of the dilaton parameter  $r_2$ and the spin parameter $a$ within the EMDA model framework. As $r_2$ increases, it affects the energy flux, the radiation temperature, and the observed luminosity of the thin accretion disk. Spectral distributions exhibit significant differences for the Kerr-Sen BH compared to the Kerr BH, providing valuable insights through observations.

Furthermore, we investigated the shadow of Kerr-Sen BH in a uniform plasma background. We computed the geodesic equations for photons and determined the shadow's boundary using celestial coordinates. Our findings indicate that, for an observer angle $\theta=90^\circ$, the BH's shadow increases as the homogeneous plasma parameter $k$ increases, with fixed parameters $r_2$ and $a$. Conversely, when $k$ and $a$ are fixed, the shadow decreases with increasing dilaton parameter.

Finally, using observational data from M87* and Sgr A*, we constrained the model parameters within $1\sigma$ and $2\sigma$ intervals based on the diameter $d_{sh}$ of shadow, enhancing our understanding of BH properties.

\begin{acknowledgments}
This work was supported by the National Key R\&D Program of China (Grants No. 2022YFA1403700),  NSFC (Grants No. 12141402, 12334002, 12333008), the SUSTech-NUS Joint Research Program, Center for Computational Science and Engineering at Southern University of Science and Technology, and Hebei Provincial Natural Science Foundation of China (Grant No. A2021201034).

\end{acknowledgments}

\appendix
\bibliographystyle{unsrt}
\bibliography{xijipan}

\end{document}